\documentclass[twoside,11pt]{article}

\pdfoutput=1

\usepackage{subcaption} 
\usepackage{jmlr2e}
\usepackage{color}
\usepackage{amsmath}

\jmlrheading{1}{2017}{1-48}{4/00}{10/00}{meila00a}{Alan Morningstar and Roger G. Melko}
\ShortHeadings{Deep Learning the Ising Model Near Criticality}{Morningstar and Melko}
\firstpageno{1}

\begin{document}

\title{Deep Learning the Ising Model Near Criticality}

\author{\name Alan Morningstar \email aormorningstar@gmail.com \\
       \name Roger G. Melko \email rgmelko@uwaterloo.ca \\
       \addr Perimeter Institute for Theoretical Physics\\
       Waterloo, Ontario, N2L 2Y5, Canada\\
       and\\ 
       Department of Physics and Astronomy
       \\University of Waterloo
       \\Waterloo, Ontario, N2L 3G1, Canada}

\editor{*}

\maketitle

\begin{abstract}
It is well established that neural networks with deep architectures perform better than shallow networks for many tasks in machine learning.  In statistical physics, while there has been recent interest in representing physical data with generative modelling, the focus has been on shallow neural networks.  A natural question to ask is whether deep neural networks hold any advantage over shallow networks in representing such data.  We investigate this question by using unsupervised, generative graphical models to learn the probability distribution of a two-dimensional Ising system.  Deep Boltzmann machines, deep belief networks, and deep restricted Boltzmann networks are trained on thermal spin configurations from this system, and compared to the shallow architecture of the restricted Boltzmann machine.  We benchmark the models, focussing on the accuracy of generating energetic observables near the phase transition, where these quantities are most difficult to approximate.  Interestingly, after training the generative networks, we observe that the accuracy essentially depends only on the number of neurons in the first hidden layer of the network, and not on other model details such as network depth or model type.  This is evidence that shallow networks are more efficient than deep networks at representing physical probability distributions associated with Ising systems near criticality.

\end{abstract}


\section{Introduction}

It is empirically well supported that neural networks with deep architectures perform better than shallow networks for certain machine learning tasks involving data with complex, hierarchal structures \citep{Hinton-Salakhutdinov2006}.  
This has led to an intense focus on developing deep neural networks for use on data sets related to natural images, speech,  video, social networks, etc.~with demonstrable success.  In regard to theoretical understanding, among the most profound questions facing the modern machine learning community is why, and under what conditions, are deep neural networks superior to shallow (single-hidden-layer) models.
Further, does the success of deep architectures in extracting features from conventional big data translate into arenas with other highly complex data sets, such as those encountered in the physical sciences?

Very recently, the statistical physics community has become engaged in exploring the representational power of generative neural networks such as the restricted Boltzmann machine (RBM; Carleo and Troyer, 2017; Wetzel, 2017)\nocite{Carleo-Troyer2017,Wetzel2017}.  There, the ``curse of dimensionality'' is manifest in the size of state space \citep{Carrrasquilla-Melko2017}, which for example, grows as $2^N$ for a system of $N$ binary (or {\it Ising}) variables.  The central problem in much of statistical physics is related to determining and sampling the probability distribution of this state space.  Therefore, the possibility of efficiently modelling a physical probability distribution with a generative neural network has broad implications for condensed matter, materials physics, quantum computing, and other areas of the physical sciences involving the statistical mechanics of $N$-body systems.

The universal approximation theorem, while establishing the capability of neural networks as generative models in theory, only requires the very weak assumption of exponentially large resources \citep{LeRoux-Bengio2008,Montufar-Morton2015}.  In both quantum and classical statistical physics applications, the goal is to calculate macroscopic {\it features} (for example, a heat capacity or a susceptibility) from the $N$-body equations.  For this, an {\it efficient} representation is desired that demands computer resources which scale polynomially, typically in the number of particles or lattice sites, $\mathcal{O}(N^c)$ where $c$ is a constant.  In this case, ``resources'' could refer to memory or time; for example, the number of model parameters, the size of the data set for training, or the number of samples required of the model to accurately reproduce physical features.  

Thus we can refine our question: are deep neural networks more efficient than shallow networks when modelling physical probability distributions, and if so, why?  
In this paper we address the dependence of generative modelling on network depth directly for the prototypical physical system of the two-dimensional (2D) lattice Ising model.  The most important feature of this system is the existence of a phase transition (or {\it critical point}), with subtle physical features that are theoretically known to require a non-trivial multi-scale description.  We define criteria for the accuracy of a generative model, based on the convergence of the energy and energy fluctuations of samples generated from it, to the exact physical values calculated near the phase transition.  Using various types of generative neural networks with a multitude of widths and depths, we find the surprising result that accurate modelling depends only on the {\it architecture} of the network---defined as the set of integers specifying the number of neurons in each network layer.  Furthermore, we show that the accuracy of the neural network only depends significantly on the number of hidden units in the first hidden layer.  This illustrates that in this physical example, depth does not contribute to the representational efficiency of a generative neural network, even near a non-trivial scale-invariant phase transition.

\section{The 2D Ising System near Criticality}

In order to benchmark the representational power of generative neural networks, we replace naturally occurring data common to machine learning applications with synthetic data for a system intimately familiar to statistical physicists, the 2D Ising model.  This system is defined on an $N$-site square lattice, with binary variables (spins) $x_i \in \{ 0, 1 \}$ at each lattice site $i$.  The physical probability distribution defining the Ising model is the Boltzmann distribution
\begin{equation}
q(x) = \frac{1}{Z} \exp(- \beta H(x)) \label{eqn:BoltzD}
\end{equation}
determined by the Hamiltonian (or physical energy) $H(x) = - \sum_{\langle i,j \rangle} \sigma_i \sigma_j$, where $\sigma_i = 2x_i - 1$.  Here, $Z$ is the normalization factor (or partition function), and $\langle i, j \rangle$ denotes nearest-neighbour sites.  As a function of inverse temperature $\beta = 1/k_B T$ (with $k_B = 1$), the Ising system displays two distinct phases, with a phase transition at $T_c = 2/\log \big( 1 + \sqrt{2} \big) \approx 2.2693$ in the thermodynamic limit where $N \rightarrow \infty$ \citep{Kramers41,Onsager44}.  In this limit, the length scale governing fluctuations---called the {\it correlation length $\xi$}---diverges to infinity resulting in emergent macroscopic behavior such as singularities in measured observables like heat capacity or magnetic susceptibility.  Phase transitions with a diverging $\xi$ are called {\it critical points}.  Criticality makes understanding macroscopic phenomena (or features) from the microscopic interactions encoded in $H(x)$ highly non-trivial.  The most successful theoretical approach is the renormalization group (RG), which is a muli-scale description where features are systematically examined at successively deeper coarse-grained levels of scale \citep{WilsonFisher}.  Physicists have begun to explore in ernest the conceptual connection between deep learning and the RG \citep{Mehta-Schwab2014,Koch-Janusz-Ringel2017}.
\begin{figure}
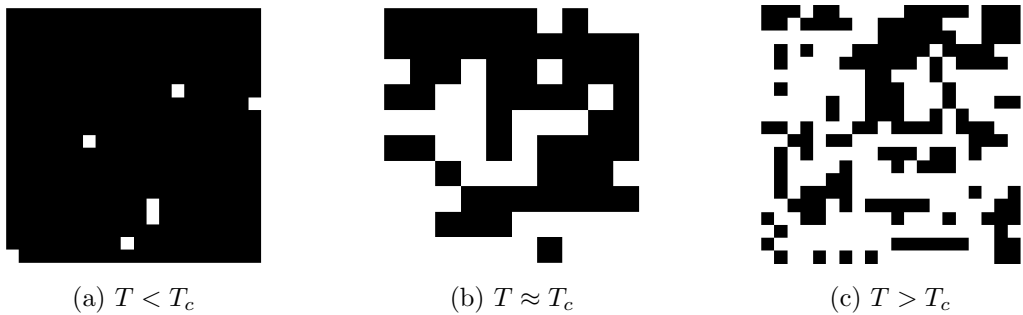

\centering
\begin{subfigure}[t]{.33\textwidth}
  \centering
  \includegraphics[width=0.7\columnwidth]{{{2DIsing_TypicalState_N=400_T=1.8}}}
  \caption{$T<T_c$}
\end{subfigure}%
\begin{subfigure}[t]{.33\textwidth}
  \centering
  \includegraphics[width=0.7\columnwidth]{{{2DIsing_TypicalState_N=400_T=2.8}}}
  \caption{$T \approx T_c$}
\end{subfigure}%
\begin{subfigure}[t]{.33\textwidth}
  \centering
  \includegraphics[width=0.7\columnwidth]{{{2DIsing_TypicalState_N=400_T=3.8}}}
  \caption{$T>T_c$}
\end{subfigure}
\caption{\label{fig:Ising_spins} Representative Ising spin configurations on a $N=20 \times 20$ site lattice.  In (a) is the ferromagnetic phase below $T_c$; (b) is the critical point or phase transition; and (c) is above $T_c$ in the paramagnetic phase.  Such configurations play the role of 2D images for the training of generative neural networks in this paper.}
\end{figure}

We concentrate on two-dimensional lattices with finite $N$, where remnants of the true thermodynamic ($N \rightarrow \infty$) phase transition are manifest in physical quantities.  We are interested in approximately modelling the physical distribution $q(x)$, both at and near this phase transition, using generative neural networks.  Real-space spin configurations at fixed temperatures play the role of data sets for training and testing the generative models of the next section.  Configurations---such as those illustrated in Figure~\ref{fig:Ising_spins}---are sampled from Equation~\ref{eqn:BoltzD} using standard Markov Chain Monte Carlo techniques described elsewhere; see for example,~\citet{Newman-Barkema1999}.  For a given temperature $T$ and lattice size $N$, training and testing sets of arbitrary size can be produced with relative ease, allowing us to explore the representational power of different generative models without the concern of regularization due to limited data.

In order to quantify the representational power of a given neural network, we use the standard recently introduced by \citet{Torlai-Melko2016}, which is the faithful recreation of physical observables from samples of the visible layer of a fully-trained generative neural network.  As a reference, observables such as average energy $E$ and heat capacity $C = \partial E / \partial T$ are calculated from the physical distribution, that is, $\langle E \rangle = Z^{-1} \sum_x q(x) H(x)$ and $\langle C \rangle = (\langle E^2 \rangle - \langle E \rangle^2)/T^2$. Using shallow RBMs, Torlai and Melko examined the convergence of these and other physical observables for 2D Ising systems on different size lattices.  For a lattice of $N$ Ising spins, the number of hidden neurons in the RBM, $N_{h_1}$ (see below), was increased until the observables modelled by the RBM approached the exact, physical values calculated by Monte Carlo.  In the RBM, $N$ also represents the number of visible units.  Thus, the ratio $\alpha \equiv N_{h_1}/N$---describing the network resources per lattice site---serves as a convergence parameter measuring the overall efficiency of this machine learning approach. It was shown by \citet{Chen-Xiang2017} that this ratio is bounded above by $\alpha = 2$ for the same 2D Ising model we consider in this paper, a bound consistent with the results of numerical simulations detailed in sections below.

\begin{figure}
	\centering
	\begin{subfigure}[t]{.5\textwidth}
		\centering
		\includegraphics[width=0.85\columnwidth]{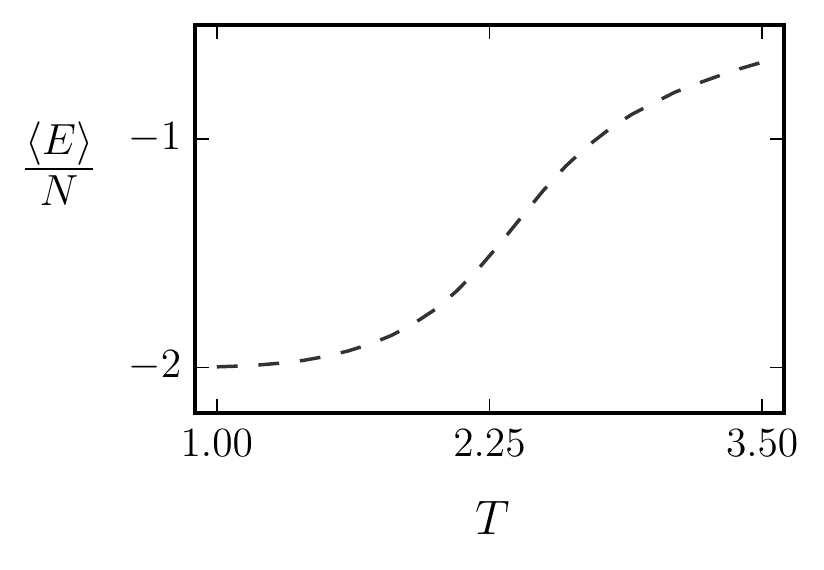}
		\caption{Energy per Ising spin.}
	\end{subfigure}%
	\begin{subfigure}[t]{.5\textwidth}
		\centering
		\includegraphics[width=0.85\columnwidth]{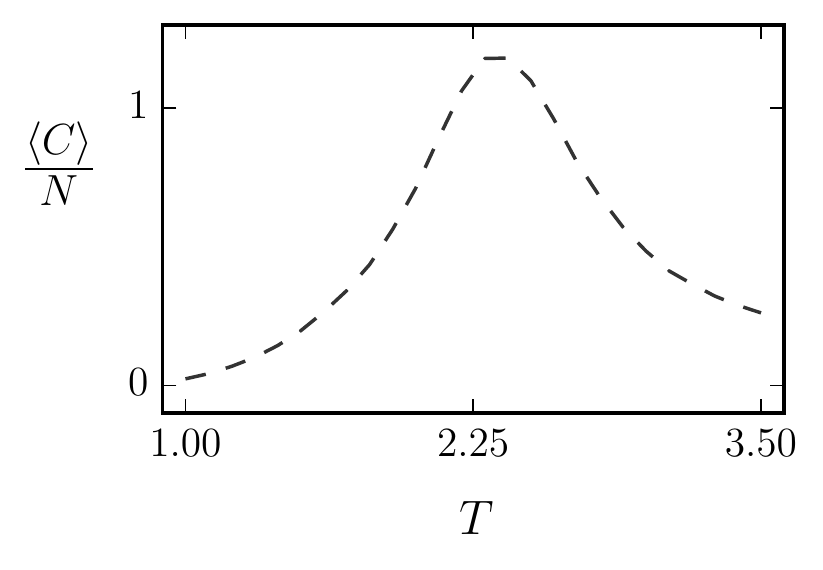}
  		\caption{Heat capacity per Ising spin.}
	\end{subfigure}%
\caption{\label{fig:MC64} Physical data for the energy and heat capacity of an $N=64$ square-lattice Ising system, calculated using Markov Chain Monte Carlo.}
\end{figure}

In Figure~\ref{fig:MC64} we illustrate the energy $E$ and heat capacity $C$ calculated via Monte Carlo for the $N = 64$ square-lattice Ising model.  Near the critical point $T_c  \approx 2.2693$, $C$ retains a finite-size remnant of the divergence expected when $N \rightarrow \infty$.  As observed by Torlai, it is here where the accuracy of the modelled neural network data suffers most from an insufficient number of hidden units.  In this paper, we therefore concentrate on the convergence of $E$ and $C$ near $T_c$ as a function of model parameters.

\section{Generative Graphical Models}

In this section, we briefly introduce the shallow and deep stochastic neural networks that serve as generative graphical models for reproducing Ising observables.  These stochastic neural networks are made up of a set of binary neurons (nodes) which can be in the state $0$ or $1$, each with an associated bias, and weighted connections (edges) to other neurons.  For this type of model, neurons are partitioned into two classes, visible and hidden neurons, such that the model defines a joint probability distribution $p(v,h)$ over the visible and hidden neurons, $v$ and $h$ respectively (or $h_1$, $h_2$, and so on for multiple hidden layers).  The weights and biases of a model---collectively referred to as ``model parameters'', and indicated by $\theta$---determine the state of each neuron given the state of other connected neurons.  They therefore define the model and parametrize the probability distribution $p (v,h)$ given a fixed architecture, which we define to specify only the number of neurons in each layer that makes up the network.  A model's parameters can be trained such that its distribution over the visible units matches a desired probability distribution $q(v)$, for example, the thermal distribution of Ising variables ($v = x$) in Equation~\ref{eqn:BoltzD}. More explicitly, given data samples from $q(v)$, $\theta$ are adjusted so that
\begin{eqnarray}
p(v) \equiv \sum_{\{h\}} p(v,h) \approx q(v).\nonumber
\end{eqnarray}
Once training is complete, $\theta$ are fixed, and the states of both visible and hidden neurons can be importance sampled with block Gibbs sampling, as long as the graph is bipartite.  Since the purpose of the model distribution is a faithful representation of the physical distribution, this allows us to calculate physical estimators from the neural network, for example, $\langle E \rangle = \| \mathcal{S} \|^{-1} \sum_{v \in \mathcal{S}} H(v)$, where $\mathcal{S}$ is a set of configurations importance sampled from the network's visible nodes.  It is the convergence of these estimators---in particular, $\langle E \rangle$ and $\langle C \rangle$---to the true values obtained via Monte Carlo that we will focus on in later sections.

Training such generative models requires data, which in this paper is the binary spin configurations of an $N$-site 2D lattice Ising model.  Training often presents practical limitations that dictate which types of stochastic neural network are effective.  The major practical limitation is the problem of inferring the state of the hidden neurons given the state of the visible layer, which is often clamped to values from the data during training.  Therefore, we only use models in which such inference can be done efficiently.  Below we outline these models, which we compare and contrast for the task of modelling a physical distribution in Section~\ref{sec:results}.

\subsection{Restricted Boltzmann Machines \label{subsec:RBM}}

\begin{figure}
\centering
\includegraphics[width=0.3\columnwidth]{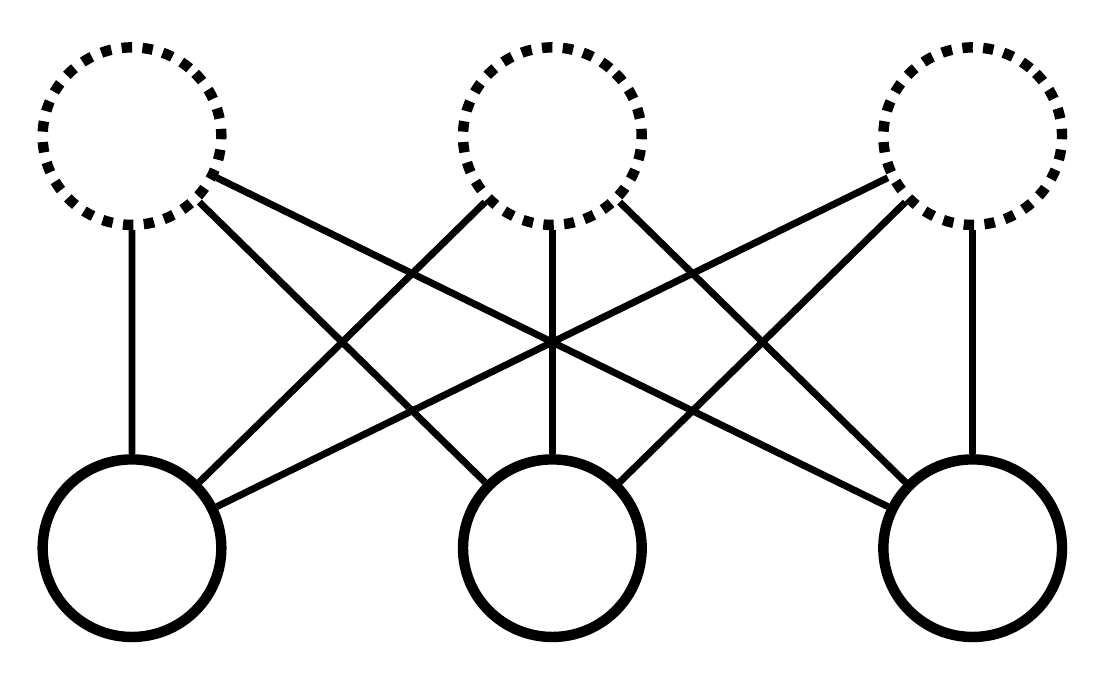}
\caption{\label{fig:RBMgraph} The restricted Boltzmann machine graph.  The solid circles at bottom represent visible nodes, $v$.  The dashed circles at top represent hidden nodes, $h$.  Straight lines are the weights $W$.}
\end{figure}

When visible and hidden nodes are connected in an undirected graph by edges (representing a weight matrix), the model is called a {\it Boltzmann machine} \citep{Ackley1985}.  The original Boltzmann machine with all-to-all connectivity suffers from intractable inference problems.  When the architecture is restricted so that nodes are arranged in layers of visible and hidden units, with no connection between nodes in the same layer, this problem is alleviated.  The resulting model, which is depicted in Figure~\ref{fig:RBMgraph}, is known as an RBM \citep{Smolensky1986}.
The joint probability distribution defining the RBM is
\begin{eqnarray}
p(v,h) = \frac{\exp \left( b^T v + c^T h + v^T W h \right)}{Z} \nonumber,
\end{eqnarray}
where $Z$ is a normalization constant or partition function, $b$ and $c$ are biases for the visible and hidden neurons respectively, and $W$ is the matrix of weights associated to the connections between visible and hidden neurons.  This is known as an energy-based model, because it follows the Boltzmann distribution for the ``energy'' function
\begin{eqnarray}
	E_{\mathrm{RBM}}(v,h) = -b^T v - c^T h - v^T W h.\nonumber
\end{eqnarray}

In order to train the RBM \citep{Hinton2012}, the negative log-likelihood $\mathcal{L}$ of the training data set $\mathcal{D}$ is minimized over the space of model parameters $\theta$, where $\mathcal{L}(\mathcal{D}) = - \frac{1}{\| \mathcal{D} \|} \sum_{v \in \mathcal{D}} \log p(v)$. In other words, the model parameters of the RBM are tuned such that the probability of generating a data set statistically similar to the training data set is maximized.  This is done via stochastic gradient descent, where instead of summing over $v\in \mathcal{D}$, the negative log-likelihood of a mini-batch $\mathcal{B}$ (with $\| \mathcal{B} \| \ll \| \mathcal{D} \|$) is minimized. The mini-batch is changed every update step, and running through all $\mathcal{B} \in \mathcal{D}$ constitutes one epoch of training.  The gradient of the negative log-likelihood must be calculated with respect to weights and biases, and consists of two terms as described in Appendix A; the data-dependent correlations, and the model-dependent correlations.  The latter term is computed approximately by running a Markov chain through a finite number $k$ steps starting from the mini-batch data.  This approximation is called {\it contrastive divergence} (CD-$k$, Hinton, 2002)\nocite{Hinton2002} which we use to train the models in this paper.  In order to use the CD-$k$ learning algorithm, efficient block Gibbs sampling must be possible.  Because of the restricted nature of the RBM, the hidden variables are conditionally independent given the visible variables, and the posterior factorizes.  The state of the hidden units can then be exactly inferred from a known visible configuration in the RBM, using the conditional probability
\begin{equation}
p(h_i=1|v)=\sigma \left( \left( c^T+v^T W \right)_i \right), \label{eqn:RBMInference}
\end{equation}
with a similar expression for $p(v_i=1|h)$.
For reference, we also provide the derivation of this conditional probability distribution in Appendix A.

Once trained, samples of the RBM's probability distribution can be generated by initializing $v$ to random values, then running Gibbs sampling as a Markov Chain ($v_0 \mapsto h_0 \mapsto v_1 \mapsto h_1 \cdots $ ).  Focussing on the visible configurations thus produced, like standard Monte Carlo after a sufficient warm-up (or equilibration) period, they will converge to importance samples of $p(v)$, allowing for the approximation of physical estimators, such as $\langle E \rangle$ or $\langle C \rangle$.  Before showing results for these, we further discuss deep generalizations of the RBM in the next sections.

\subsection{Deep Boltzmann Machine \label{subsec:DBM}}
\begin{figure}
\centering
\begin{subfigure}[t]{.32\textwidth}
  \centering
  \includegraphics[width=0.85\columnwidth]{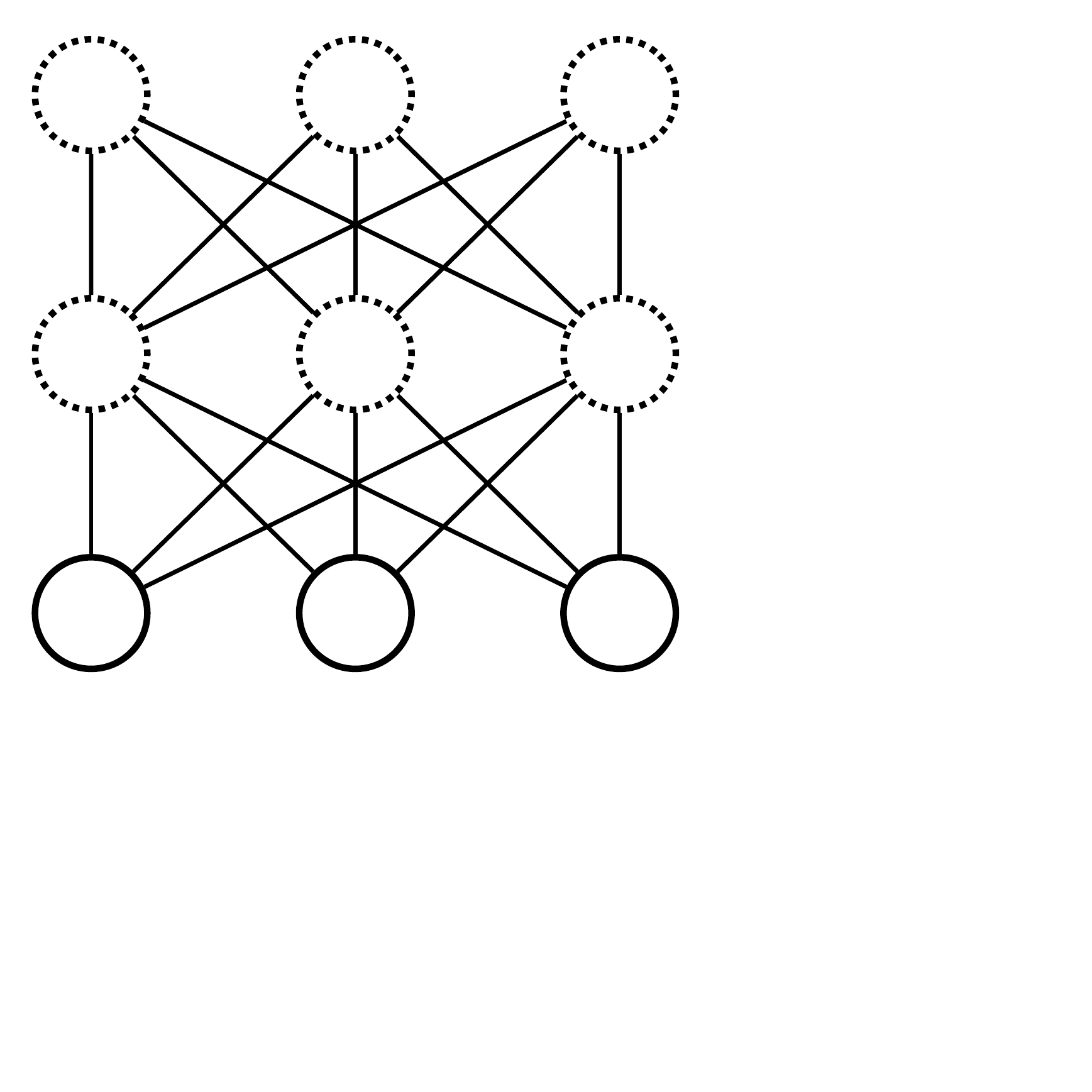}
  \caption{DBM}
\end{subfigure}%
\begin{subfigure}[t]{.32\textwidth}
  \centering
  \includegraphics[width=0.85\columnwidth]{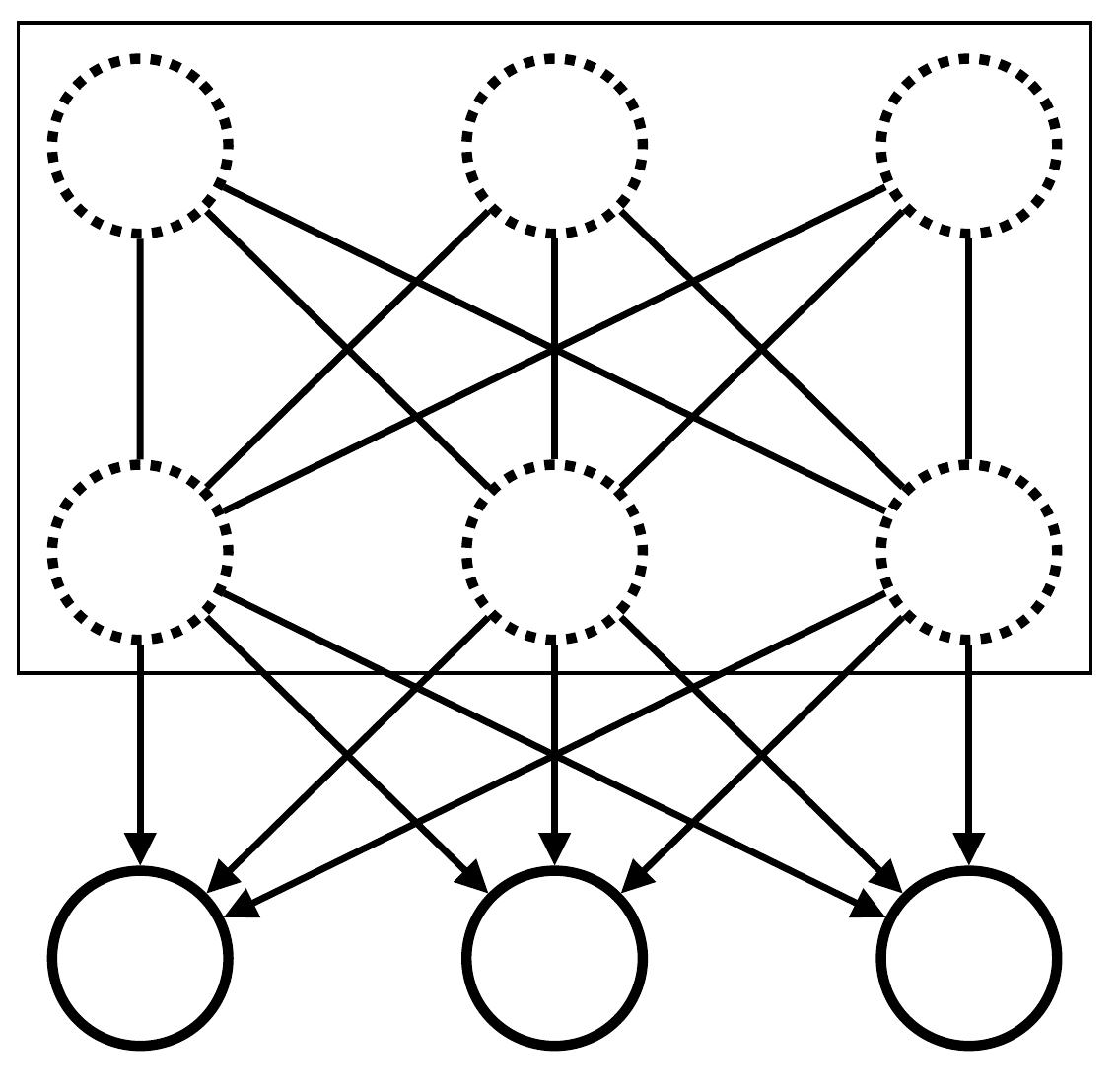}
  \caption{DBN}
\end{subfigure}%
\begin{subfigure}[t]{.32\textwidth}
  \centering
  \includegraphics[width=0.85\columnwidth]{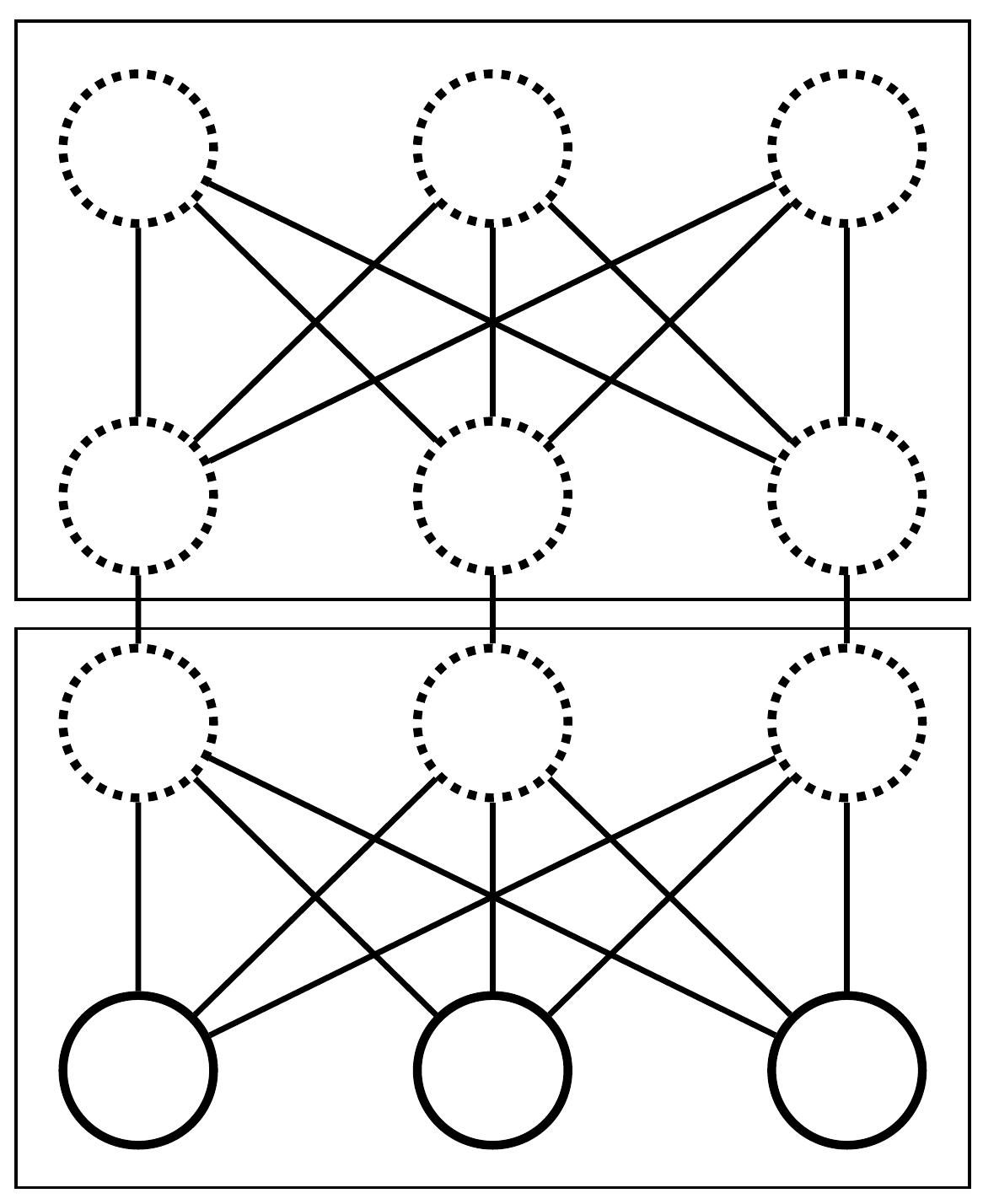}
  \caption{DRBN}
\end{subfigure}%
\caption{\label{fig:deepGraphs} The deep generalizations of the restricted Boltzmann machine.  Dashed circles represent hidden nodes, with $h_1$ in the layer above $v$, $h_2$ in the layer above $h_1$, and so on.  Directed (undirected) edges depict a one-way (two-way) dependence of the neurons they connect.  A closed rectangle indicates a constituent restricted Boltzmann machine.}
\end{figure}

A {\it deep Boltzmann machine} (DBM; Salakhutdinov and Hinton, 2009)\nocite{Salakhutdinov-Hinton2009} has one visible and two or more hidden layers of neurons, which we label $v, h_1, h_2, \cdots$ (Figure \ref{fig:deepGraphs}). 
It is a deep generalization of the restricted Boltzmann machine, and is similarly an energy-based model, with the energy for the two-hidden-layer case being
\begin{eqnarray}
E_{\mathrm{DBM}}(v,h_1,h_2) &=& -b^T v - c^T h_1 - d^T h_2 - v^T W_0 h_1 - h_1^T W_1 h_2.\nonumber
\end{eqnarray}
Sampling $p(h_1,h_2|v)$ cannot be done exactly as in the case of the RBM. This is because the conditional probabilities of the neurons are
\begin{eqnarray}
p(v_i=1|h_1) &=& \sigma\left( \left( b^T+W_0 h_1 \right)_i \right) \nonumber \\
p((h_1)_i=1|v,h_2) &=& \sigma\left( \left( c^T+v^T W_0 + W_1 h_2 \right)_i\right) \nonumber \\
p((h_2)_i=1|h_1) &=& \sigma\left( \left(d^T+h_1^T W_1 \right)_i \right).\nonumber 
\end{eqnarray}
However, notice that the even and odd layers are conditionally independent. That is, for a three layer DBM, $h_1$ may be sampled given knowledge of $\{v,h_2\}$, and $\{v,h_2\}$ may be sampled given the state of $h_1$. Therefore, in order to approximately sample $p(h_1,h_2|v_0)$ for some $v_0$, successive Gibbs sampling of even and odd layers with the visible neurons clamped to $v_0$ is performed. Under such conditions, the state of the hidden neurons will equilibrate to samples of $p(h_1,h_2|v_0)$. This is how the inference $v_0 \mapsto h_1,h_2$ is performed in our training procedure. In order to generate model-dependent correlations, the visible neurons are unclamped and Gibbs sampling proceeds for $k$ steps. The parameter updates for the DBM are similar to the CD-$k$ algorithm for the RBM described in Section \ref{subsec:RBM} and explicitly given in Appendix A \citep{Salakhutdinov-Hinton2009}.

Pre-training the DBM provides a reasonable initialization of the weights and biases by training a stack of RBM's and feeding their weights and biases into the DBM.  Our pre-training procedure was inspired by \citet{Hinton-Salakhutdinov2012}, however it varies slightly from their approach because the two-layer networks used in our numerical results below are small enough to be trained fully without ideal pre-training.  Our procedure goes as follows for a two-layer DBM.  First, an RBM is trained on the training data set $\mathcal{D}$, and is then used to infer a hidden data set $\mathcal{H}$ via sampling the distribution $p(h|v\in \mathcal{D})$.  A second RBM is then trained on $\mathcal{H}$.  Weights from the first and second RBMs initialize the weights in the first and second layers of the DBM respectively. DBM biases are similarly initialized from the RBM stack, however in the case where the biases in the first hidden layer of the DBM can be taken from either the hidden layer in the first RBM, or the visible layer in the second RBM, these two options are averaged.  We emphasize that this procedure is not ideal for DBMs with more than three layers.

Once fully trained, the probability distribution associated to the DBM can be sampled, similarly to the RBM, by initializing all neurons randomly, then running the Markov chain even layers $\mapsto$ odd layers $\mapsto$ even layers $\cdots$ etc.~until the samples generated on the visible neurons are uncorrelated with the initial conditions of the Markov chain.

\subsection{Deep Belief Network}
The {\it deep belief network} (DBN) of Figure \ref{fig:deepGraphs} is another deep generalization of the RBM. It is not a true energy-based model, and only the deepest layer of connections are undirected. The deepest two hidden layers of the DBN form an RBM, with the layers connecting the hidden RBM to the visible units forming a feed forward neural network. The DBN is trained by first performing layer-wise training of a stack of RBMs as detailed by \citet{Hinton-Osindero2006}, and similarly to the aforementioned pre-training procedure in Section~\ref{subsec:DBM}. The model parameters of the deepest hidden RBM are used to initialize the deepest two layers of the DBN. The remaining layers of the DBN take their weights and biases from the weights and visible biases of the corresponding RBMs in the pre-training stack.  Note, there exists a fine tuning procedure called {\it the wake-sleep algorithm} \citep{Hinton-Neal1995}, which further adjusts the model parameters of the DBN in a non-layer-wise fashion. However, performing fine tuning of the DBN was found to have no influence on the results described below.

Once trained, a DBN can be used to generate samples by initializing the hidden RBM randomly, running Gibbs sampling in the hidden RBM until convergence to the model distribution is achieved, then propagating the configuration of the RBM neurons downwards to the visible neurons of the DBN using the same conditional probabilities used in an RBM, Equation~\ref{eqn:RBMInference}.

\subsection{Deep Restricted Boltzmann Network}
The {\it deep restricted Boltzmann network} (DRBN) of Figure \ref{fig:deepGraphs} is a simple, deep generalization of the RBM which, like the DBN, is also not a true energy-based model \citep{Hu-Ma2016}. Inference in the DRBN is exactly the same as in the RBM, but with many layers,
\begin{eqnarray}
p((x_l)_i | x_{l-1}) &=& \sigma \left( \left(b_l^T+x_{l-1}^T W_{l-1} \right)_i \right) \nonumber \\
p((x_{l-1})_i | x_{l}) &=& \sigma \left( \left(b_{l-1}^T+W_{l-1} x_{l} \right)_i \right),\nonumber 
\end{eqnarray}
where $x_0 \equiv v$, $x_1 \equiv h_1$ etc.  However, Gibbs sampling in the DRBN is performed by doing a complete up (down) pass, where the state of neurons in all layers of the DRBN are inferred sequentially given the state of the bottom (top) neurons. The CD-$k$ algorithm is applied to train the DRBN, as it was for the RBM, by first inferring data-dependent correlations from some initial data $v_0$, then performing $k$ steps of Gibbs sampling in order to generate the model-dependent correlations.

Once trained, the DRBN is sampled by initializing the visible neurons randomly, then performing Gibbs sampling until the visible samples converge to samples of the model dependent distribution $p(v)$.

\section{Training and Results \label{sec:results}}

In this section, we investigate the ability of the generative graphical models, introduced in the previous section, to represent classical statistical mechanical probability distributions, with a particular interest in the comparison between deep and shallow models near the phase transition of the two-dimensional Ising system.

In order to produce training data, we use standard Markov chain Monte Carlo techniques with a combination of single-site Metropolis and Wolff cluster updates, where one Monte Carlo step consists of $N$ single-site updates and one cluster update, and $N$ is the number of sites on the lattice.  Importance sampling thus obtained $10^5$ independent spin configurations for each $T\in [1.0,3.5]$ in steps of $\Delta T=0.1$.  An equilibration time of $N^3$ Monte Carlo steps and a decorrelation time of $N$ steps were used, where in this paper we concentrate on $N=64$ only.

Using these data sets, we trained shallow (RBM) and deep (DBM, DBN, DRBN) generative models to learn the underlying probability distribution of the Monte Carlo data for each value of $T$. Similar to the work done by \citet{Torlai-Melko2016}, restricted Boltzmann machines of architecture $(N,N_{h_1})$ for $N=64$ and various $N_{h_1}$ were trained. Further, we trained two-hidden-layer deep models of architecture $(N,N_{h_1},N_{h_2})$ in order to investigate the efficiency with which deep architectures can represent the thermal distribution of Ising systems.   Networks with two hidden layers were chosen to reduce the large parameter space to be explored, while still allowing us to quantify the difference between shallow and deep models.  Training hyper-parameters for each model are given in Table~\ref{tab:hyperparameters}, and values of $N_{h_1}$ and $N_{h_2}$ used in this work were  $\{ 8,16,24,32,40,48,56,64 \}$.
\begin{table}
\begin{center}
\begin{tabular}{|c||r|r|r|r|}
\hline
hyperparameter & RBM & DBM & DBN & DRBN\\
\hline
\hline
$k$ & 10 & 10 & 10 & 5\\
\hline
equilibration steps & NA & 10 & NA & NA\\
\hline
training epochs & $4\times 10^3$ & $3\times 10^3$ & $3\times 10^3$ & $3\times 10^3$\\
\hline
learning rate & $5\times 10^{-3}$ & $10^{-3}$ & $10^{-4}$ & $10^{-4}$\\
\hline
mini-batch size & $10^2$ & $10^2$ & $10^2$ & $10^2$\\
\hline
\end{tabular}
\caption{\label{tab:hyperparameters} Values of training hyper-parameters for each model, trained with contrastive divergence CD-$k$.}
\end{center}
\end{table}

\begin{figure}
	\begin{center}
		\includegraphics[width=0.60\columnwidth]{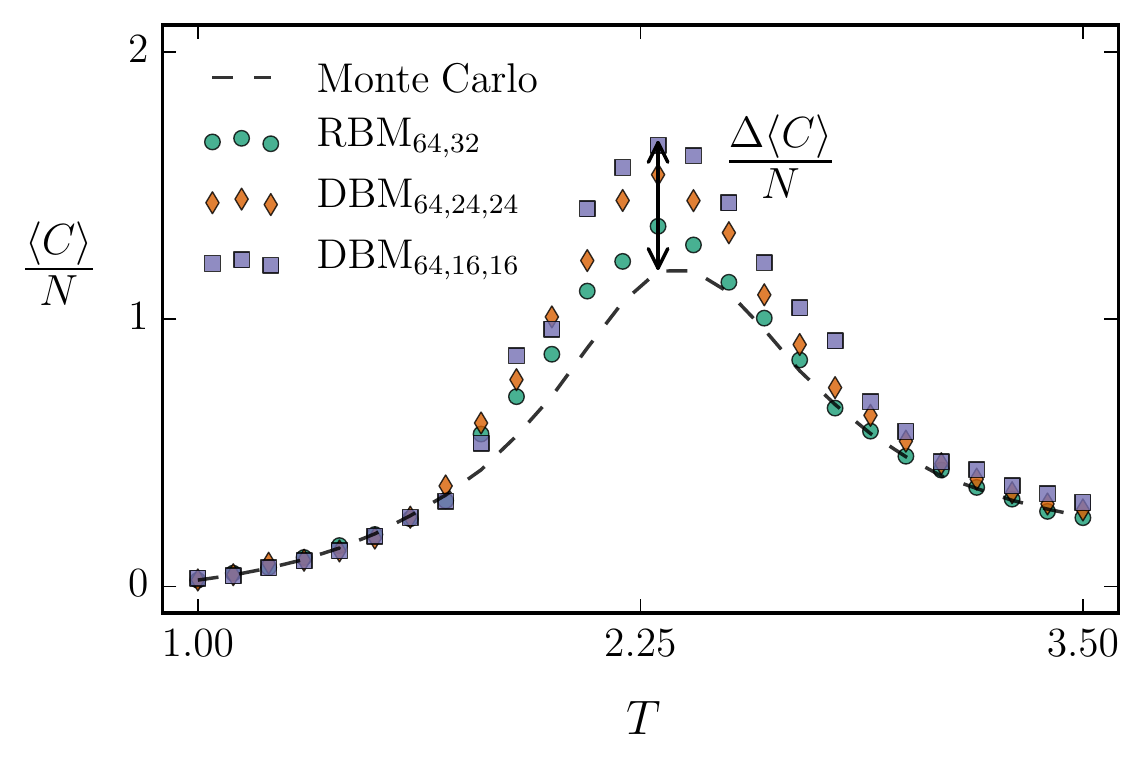}
		\caption{\label{fig:RBM_32vsDBM} Measurements of heat capacity per Ising spin on samples generated by trained shallow and deep models of architecture $(64,32)$, $(64,16,16)$, and $(64,24,24)$. Measurements are compared to Monte Carlo values of the data on which the models were trained.}
	\end{center}
\end{figure}
In order to quantify how well the models captured the target physical probability distribution, each model was used to generate a set $\mathcal{S}$ of $10^4$ spin configurations on which the observables $\langle E \rangle$ and $\langle C \rangle$ were computed and compared to the exact values from Monte Carlo simulations.

We begin by benchmarking our calculations on the case of the shallow RBM, studied previously by \citet{Torlai-Melko2016}.  Similar to the results reported in that work, we find that for all temperatures an RBM with $N_{h_1} =  N$ accurately captures the underlying Ising model distribution, a result consistent with the upper bound of $N_{h_1} \le 2N$ found by \citet{Chen-Xiang2017} for the 2D Ising model.  RBMs with less hidden neurons fail to completely capture the observables of the data, particularly near the critical region defined by $T \approx T_c$.  Furthermore, we find here that all deep models with architecture $N_{h_1}=N_{h_2}=N$ can also accurately represent the probability distribution generated by the $N=64$ Ising Hamiltonian at any temperature.  Below we concentrate on $N_{h_1},N_{h_2} < N$ in order to systematically investigate and compare the representational power of each model type. 

Figure \ref{fig:RBM_32vsDBM} shows a heat capacity calculated from the spin configurations on the visible nodes.  As an illustrative example, we present a shallow RBM with 32 hidden units, compared to two different DBMs each with two hidden layers.  To reduce parameter space, we let $N_{h_1}=N_{h_2}$ for these deep models.  From this figure it is clear that the DBM with the same number of total hidden neurons ($\mathrm{DBM}_{64,16,16}$) performs worse than the DBM with approximately the same total number of model parameters ($\mathrm{DBM}_{64,24,24}$) as the RBM.  Remarkably, both deep networks are outperformed by the shallow RBM, suggesting it is more efficient to put additional network resources into the first hidden layer, than to add more layers in a deep architecture.

\begin{figure}
	\begin{center}
		\includegraphics[width=0.6\columnwidth]{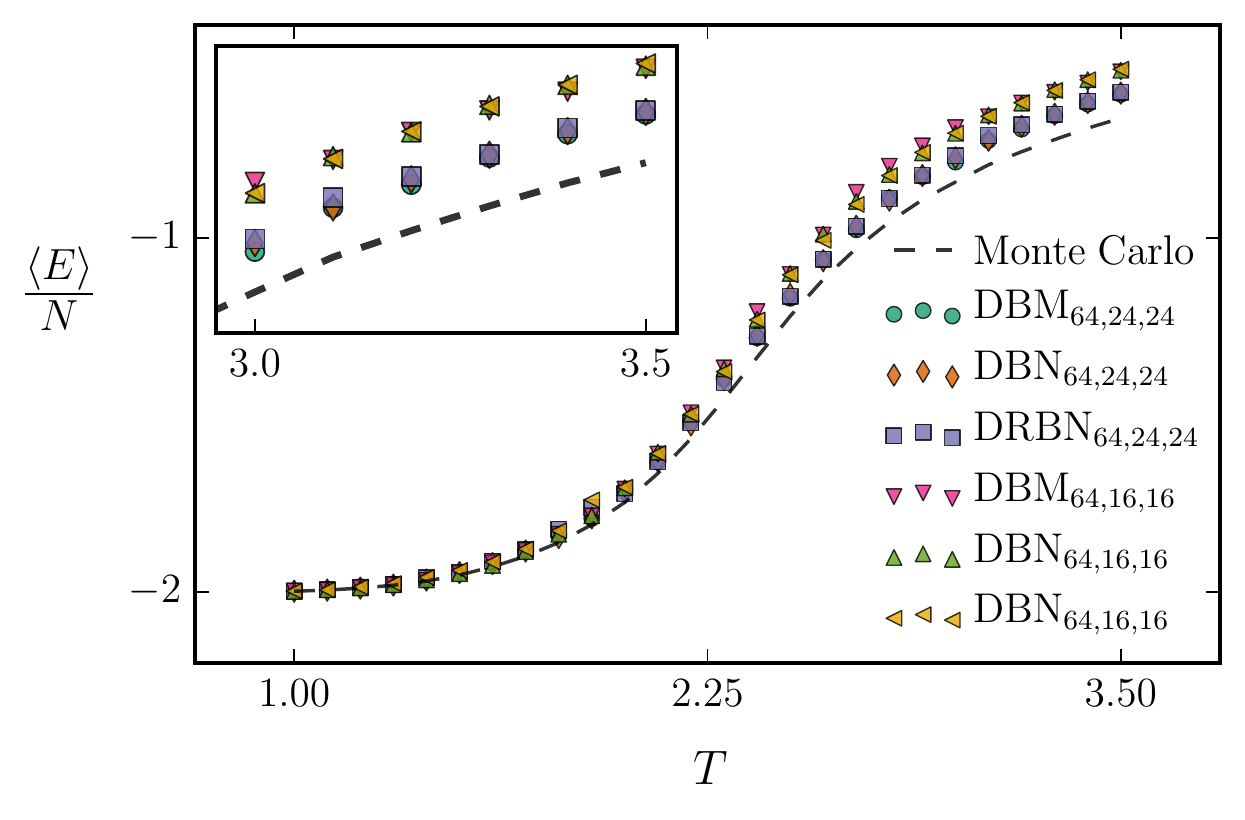}
		\caption{\label{fig:EvsT_allModels} Measurements of energy $E$ per Ising spin on samples generated by trained shallow and deep models of various architectures.  Measurements are compared to Monte Carlo values of the data on which the models were trained.}
	\end{center}
\end{figure}
In Figure \ref{fig:EvsT_allModels} we examine $\langle E \rangle$ for deep models with two different architectures, namely $N_{h_1} = N_{h_2} = 16$ and $N_{h_1} = N_{h_2} = 24$.  In this plot, two families of curves become apparent, associated with these two different network architectures.  From this, it appears that between all the different deep models, the differences in performance correspond only to the number and arrangement of hidden nodes (i.e. the ``architecture'' of the network), and not to the specific model type and training procedure, which varies considerably between DBM, DBN, and DRBN.

\begin{figure}
\centering
\includegraphics[width=0.6\columnwidth]{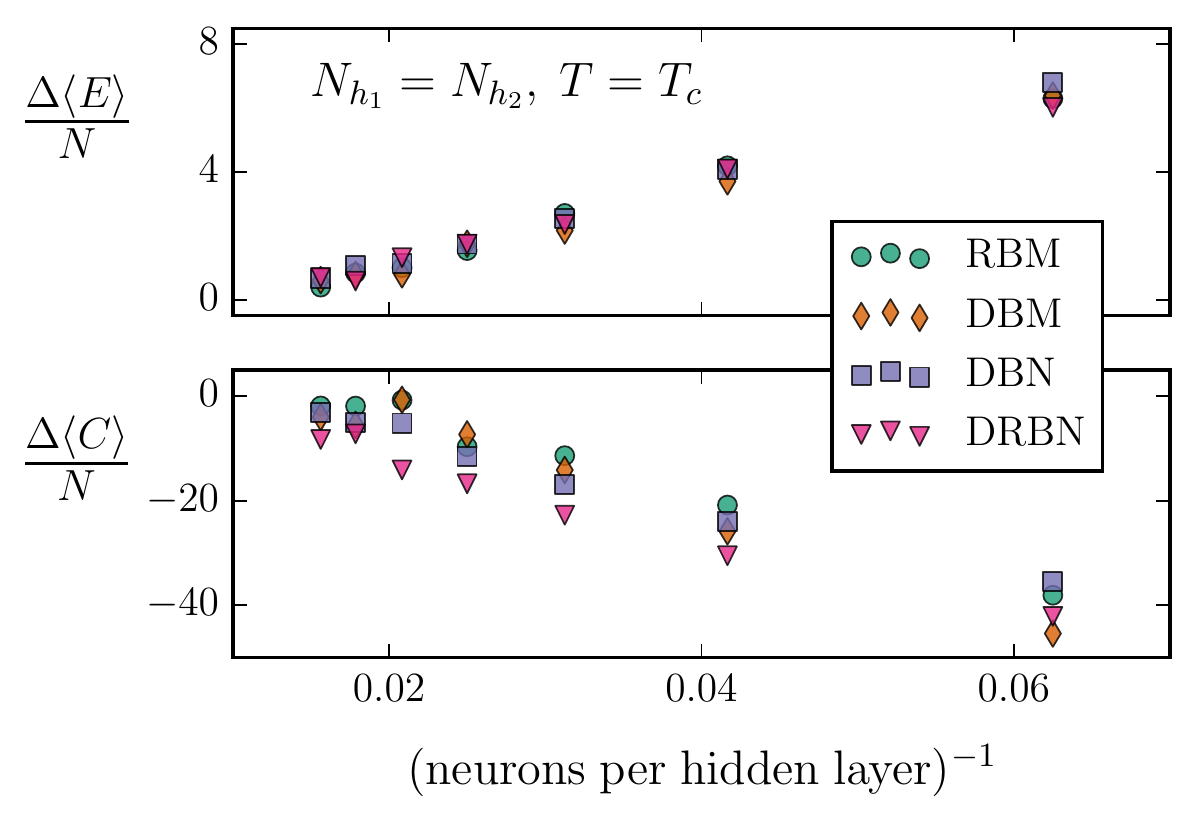}
\caption{\label{fig:percentError} The error in reproducing the energy and heat capacity per Ising spin of the training data as a function of the number of hidden neurons per layer for all model types.}
\end{figure}

To further investigate this point, we next compare RBMs with deep models having the same number of hidden units in each hidden layer ($N_{h_1} = N_{h_2}$) as the RBMs do in $h_1$.  Figure~\ref{fig:percentError} compares the percent error at $T=T_c$ in reproducing the physical observables energy and heat capacity.  In this figure, we see the expected monotonic trends of both $\Delta \langle E \rangle$ and $\Delta \langle C \rangle$, which are defined as the difference between the generated quantity and its exact value at the critical point (see Figure~\ref{fig:RBM_32vsDBM}).  This demonstrates that, for all models, the accuracy of reconstructing physical observables increases with the number of neurons per hidden layer.  More surprisingly however, since values of $\Delta \langle E \rangle$ and $\Delta \langle C \rangle$ are not significantly different across model types with equal architecture, it can also be said that adding a second hidden layer to an RBM does not improve its ability to generate accurate observables.  This is the case even when the deep networks have greater total network resources (neurons or model parameters) than the shallow RBMs, as they do in Figure~\ref{fig:percentError}.  This demonstrates again that the most efficient way to increase representational power in this case is not to add a second hidden layer, thereby increasing the depth of the neural network, but to increase the size of the only hidden layer in a shallow network.

\begin{figure}
	\begin{center}
		\includegraphics[width=1.0\columnwidth , trim=-4.5cm 0 0 0]{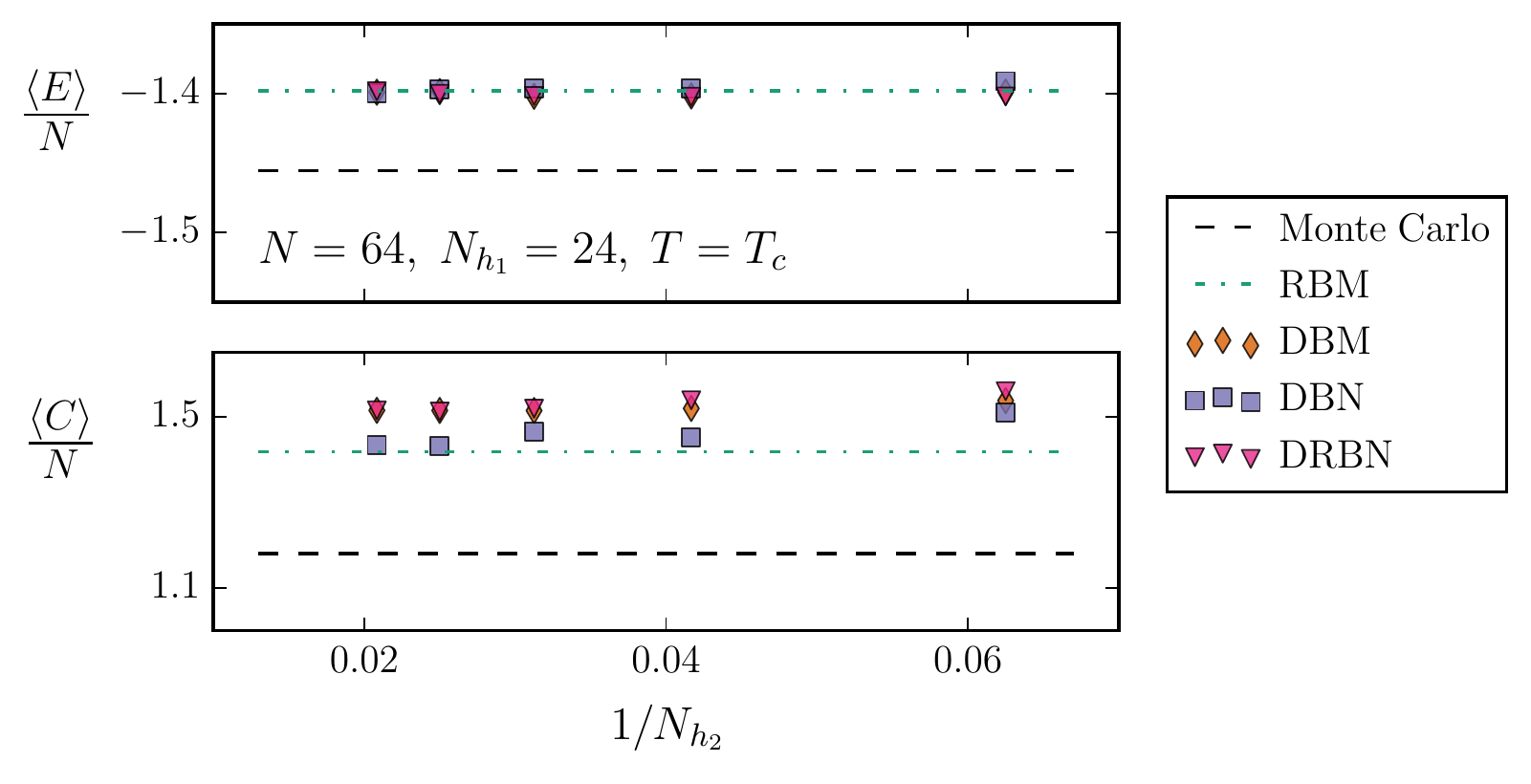}
		\caption{\label{fig:CvsnH2_EvsnH2} Energy $E$ and heat capacity $C$ per Ising spin at the critical temperature as measured on samples generated from deep models with various number of neurons in the second hidden layer.  $N_{h_1}$ is fixed, and model values are compared to Monte Carlo and an RBM, which serves as the $N_{h_2}=0$ reference point.}
\end{center}
\end{figure}

We confirm this in a striking way in Figure~\ref{fig:CvsnH2_EvsnH2}, where we plot both $\langle E \rangle$ and $\langle C \rangle$ at the critical temperature for samples generated from deep models of architecture $(64,24,N_{h_2})$, in comparison to an RBM with $N_{h_1}=24$ and the exact Monte Carlo values.  The fact that there is no clear systematic improvement in accuracy as $N_{h_2}$ increases confirms that the accuracy of deep neural networks for the Ising system is essentially independent of the second hidden layer.

\section{Conclusion \label{sec:Conclusions}}

We have investigated the use of generative neural networks as an approximate representation of the many-body probability distribution of the two-dimensional Ising system, using the restricted Boltzmann machine and its deep generalizations; the deep Boltzmann machine, deep belief network, and deep restricted Boltzmann network.  We find that it is possible to accurately represent the physical probability distribution, as determined by reproducing physical observables such as energy or heat capacity, with both shallow and deep generative neural networks.  However we find that the most efficient representation of the Ising system---defined to be the most accurate model given some fixed amount of network resources---is given by a shallow RBM, with only one hidden layer.

In addition, we have also found that the performance of a deep generative network is independent of what specific model was used.  Rather, the accuracy in reproducing observables is entirely dependent on the network architecture, and even more remarkably, on the number of units in the first hidden layer only.  This result is consistent with that of \citet{Raghu-Dickstein2017}, who find that trained networks are most sensitive to changes in the weights of the first hidden layer.  Additionally, in the wake of recent developments in information based deep learning by \citet{Tishby-Zaslavsky2015}, we remark that such behavior may be a signature of the information bottleneck in neural networks.

It is interesting to view our results in the context of the empirically well-supported fact that deep architectures perform better for supervised learning of complex, hierarchal data \citep{Lecun-Hinton2015}.  In statistical mechanical systems there is often a very simple pattern of local, translation-invariant couplings between degrees of freedom.  While there are multi-scale techniques in physics such as the renormalization group for analyzing physical systems, no obvious hierarchy of features exists to be leveraged by generative, deep learning algorithms. While it can be speculated that the power of deep networks also translates to the realm of unsupervised, generative models, for natural data sets commonly used by the machine learning community it is difficult to obtain quantitative measures of accuracy \citep{Hinton-Osindero2006,LeRoux-Bengio2008,Salakhutdinov-Hinton2009,Hu-Ma2016}.  Therefore, modelling physics data offers an opportunity to quantitatively measure and compare the performance of deep and shallow models, due to the existence of physical observables such as energy and heat capacity.  

However, considering our results, it is clear that there may be a fundamental difference between statistical physics data and real-world data common to machine learning practices, otherwise the established success of deep learning would more easily translate to applications of representing physics systems, such as thermal Ising distributions near criticality.  It is interesting to ask what precisely this difference is.

The lines of inquiry addressed in this paper could also be continued by extending this analysis to consider the efficient representation of quantum many-body wave functions by deep neural networks \citep{Torlai-Carleo2017}, and by investigating in more detail the nature of information flow in neural-network representations of physical probability distributions.

\vskip 0.5in

\acks{We would like to thank J. Carrasquilla, G. Torlai, B. Kulchytskyy, L. Hayward Sierens, P. Ponte, M. Beach, and A. Golubeva for many useful discussions. This research was supported by Natural Sciences and Engineering Research Council of Canada, the Canada Research Chair program, and the Perimeter Institute for Theoretical Physics. Simulations were performed on resources provided by the Shared Hierarchical Academic Research Computing Network (SHARCNET). Research at Perimeter Institute is supported through Industry Canada and by the Province of Ontario through the Ministry of Research and Innovation.}

\appendix
\section{ \label{app:A}}

\numberwithin{equation}{section}
\setcounter{equation}{0}

The gradient of the negative log-likelihood must be calculated with respect to weights and biases; for example, $\nabla_W (\mathcal{L}) = \langle vh^T \rangle_{p(h|v)} - \langle vh^T \rangle_{p(v,h)}$, where $\langle f \rangle_{P(x)}$ denotes the expectation value of $f$ with respect to the probability density $P(x)$. The first term in the gradient is referred to as the data-dependent correlations, and is tractable as long as sampling $p(h|v)$ is efficient, which it is for RBMs. However, the second term, referred to as model-dependent correlations, is ideally computed by running the Markov chain $v_0 \mapsto h_0 \mapsto v_1 \mapsto h_1 \cdots$ an infinite number of steps in order to sample the model distribution without bias due to the initial conditions $v_0$. This cannot be done exactly, and the standard approximation to estimate the model-dependent term is to run the Markov chain for only $k<\infty$ steps from an initial value of $v_0$ given by the mini-batch data---in practice, $k$ of order 1 may do.  For concreteness, the full update rules for the weights and biases of an RBM are then given by $\theta \mapsto \theta+\eta \Delta \theta$, where
\begin{eqnarray}
\Delta W &=& \| \mathcal{B} \| ^ {-1} \sum_{v_0 \in \mathcal{B}} v_0 h_0^T  -  v_k h_k^T \label{eqn:CDupdates1}\\
\Delta b &=& \| \mathcal{B} \| ^ {-1} \sum_{v_0 \in \mathcal{B}} v_0  -  v_k  \\
\Delta c &=& \| \mathcal{B} \| ^ {-1} \sum_{v_0 \in \mathcal{B}} h_0  -  h_k, \label{eqn:CDupdates3}
\end{eqnarray}
and $\eta$ is a hyper-parameter of the learning algorithm called the {\it learning rate}.

In order to evaluate the parameter updates in Equations~\ref{eqn:CDupdates1} to \ref{eqn:CDupdates3}, conditional probabilities in an RBM such as $p(h_i=1|v)$ must be known. Therefore, here we provide a derivation of the conditional probability distribution $p (h_i=1|v) = \sigma \left( \left( c^T+v^T W \right)_i \right)$ of an RBM.

Firstly, the marginal distribution $p(v)$ is given by
\begin{eqnarray}
p(v) &=& \sum_{h} p(v,h) \nonumber \\
&=& \frac{1}{Z} \exp \left( b^T v +\sum_{i} \log \left(1+ \exp\left( \left(c^T+v^T W \right)_i \right) \right) \right) \nonumber \\
&\equiv & \frac{1}{Z} \exp ( - \mathcal{F}(v) ) ,\nonumber 
\end{eqnarray}
where $\mathcal{F}(v)$ is known as the ``free energy''. One can interpret training an RBM as fitting $\mathcal{F}(v)$ to the Hamiltonian $H(v)$ of the training data set, sampled from Equation~\ref{eqn:BoltzD}, a perspective taken by previous studies in physics (see Huang and Wang, 2017)\nocite{Huang-Wang2017}. Next, using Bayes' theorem, $p(h|v) = \frac{1}{Z'} \prod_{i} \exp \left( (c^T+v^T W)_i h_i \right)$, meaning that the conditional probability factors, $p(h|v) = \prod_i p(h_i|v)$.  Finally, by requiring normalization, the conditional activation probabilities are
\begin{eqnarray}
p ( h_i = 1|v ) &=& \sigma \left( \left( c^T+v^T W \right)_i \right) \nonumber 
\end{eqnarray}
where $\sigma(x)=(1+\exp(-x))^{-1}$. The activation probabilities $p(v_i=1|h)$ are derived similarly.

\vskip 0.2in

\bibliography{deep_critical}

\end{document}